\documentclass[aps,prd,twocolumn]{revtex4}
\usepackage{graphicx}

\begin{document}

\parindent 0 mm
\parskip 2 mm

\title{Spacetime structure of self-similar spherically symmetric
perfect fluid solutions}

\author{B. J. Carr}

\affiliation{ 
Astronomy Unit, Queen Mary, Mile
End Road, London E1 4NS, UK}

\author{Carsten Gundlach} 

\affiliation{
Faculty of Mathematical Studies, University of Southampton,
Southampton SO17 1BJ, UK}

\date{4.12.02}



\begin{abstract}

We classify all spherically symmetric and homothetic spacetimes that
are allowed {\it kinematically} by constructing them from a small
number of building blocks. We then restrict attention to a particular
{\it dynamics}, namely perfect fluid matter with the scale-free
barotropic equation of state $p=\alpha \mu$ where $0\le \alpha<1$ is a
constant. We assign conformal diagrams to all solutions in the
complete classification of Carr and Coley, and so establish which of
the kinematic possibilities are realized for these dynamics. We pay
particular attention to those solutions which arise as critical
solutions during gravitational collapse.

\end{abstract}


\maketitle


\tableofcontents


\section{Introduction} 


Self-similar and spherically symmetric solutions play an important
role in many areas of physics. In Newtonian physics, using suitable space and time 
coordinates  $\vec x$ and $t$, self-similarity
is an ansatz $f(\vec x, t) = f_0(\vec x/t)$ for some field $f$ that is of interest. On the one hand, such an ansatz simplifies the field equations by
eliminating one coordinate. (In spherical symmetry, in particular, the
self-similar ansatz simplifies field equations that are partial
differential equations in $r$ and $t$ to ordinary differential
equations in one variable $r/t$.) On the other hand, self-similar
solutions are often attractors, or intermediate attractors, and are
therefore good approximations to {\it generic} solutions during some
stage of their evolution.

A classic example is the solution of the Riemann problem in
1-dimensional fluid dynamics, where at $t=0$ the fluid has one constant
velocity, density and entropy for $x<0$, and another set of constant
values for $x>0$. The resulting shock solution is self-similar in the
coordinate $x/t$. The reason for self-similarity is that the initial
data do not contain any length scale. The self-similar solution is a
good approximation over an intermediate range of scales, larger than the
molecular scale where the fluid assumption breaks down but
smaller than the scale on which the solution varies smoothly.  

In general relativity, the most common definition of self-similarity
is homotheticity: the spacetime admits a vector field $\xi^\mu$ with
the property that the Lie-derivative of the metric along this field is 
a constant times the metric:
\begin{equation}
\label{CSS}
{\cal L}_\xi g_{\mu\nu} = -2 g_{\mu\nu}.
\end{equation}
(The choice of the constant as $-2$ is a convention, the negative sign
being chosen so that $\xi^\mu$ points towards increasing curvature.)
This definition expresses the scale-invariance of the
spacetime metric, and is analogous to the Newtonian notion of
self-similarity. With spherical symmetry, it is possible to introduce
coordinates $t$ and $r$ so that all metric coefficients depend only on
$r/t$. In general relativity,
self-similarity has an important application in gravitational
collapse: during collapse whose initial data are close to the black
hole threshold, certain self-similar ``critical'' solutions act as universal
intermediate attractors

The continuous self-similarity (CSS) that we have discussed already
can be generalized to a discrete self-similarity (DSS), in which
$f(\vec x,t)$ depends arbitrarily on $\vec x/t$ but can also depend on
$\ln|t|$ in a periodic manner, with some period $\Delta$. Clearly such
solutions are scale-periodic rather than scale-invariant. Several
known critical solutions in gravitational collapse, in particular
those with scalar field matter and in vacuum gravity, are DSS rather
than CSS. In this paper we are interested in perfect fluids, where the
critical solutions are CSS, and from now on we restrict our discussion
to this case. We refer to CSS when we specifically mean the continuous
symmetry, and to self-similarity in any statements that are also true
for DSS.

We proceed in three steps. We first discuss the kinematic consequences
of CSS in spherical symmetry, generalizing the discussion in Ori and
Piran \cite{OriPiran} and Nolan \cite{Nolan}. We then discuss the
dynamics with perfect fluid matter and examine which of the kinematic
possibilities are realized dynamically. We assume the linear
barotropic equation of state $p=\alpha\mu$, which is the only one that
is compatible with exact self-similarity. We use the classification by
Carr and Coley \cite{CarrColey} of all spherically symmetric CSS
perfect fluid solutions to draw the corresponding conformal
diagrams. Finally, we look more closely at those solutions that appear
as critical solutions in gravitational collapse.


\section{Kinematics of spherically symmetric and homothetic
spacetimes}


In this section we discuss the kinematics of spherical symmetry,
without using the Einstein equations. Any spherically
symmetric spacetime is naturally the product of a two-dimensional
reduced spacetime $M^2$ crossed with the 2-sphere $S^2$. Any point in
the reduced spacetime corresponds to a 2-sphere of area $4\pi
R^2$. Here $R$ transforms as a scalar on the reduced spacetime, and
$R=0$ is a boundary of the reduced spacetime, which can either be a
regular center or a central curvature singularity. A local mass
function $m$ can be defined by $1-(2m/R)\equiv\nabla_\mu R\nabla^\mu
R$. Any point in the reduced spacetime 
where $\nabla_\mu R$ is timelike is a closed trapped surface in the
full spacetime and a surface where
$\nabla_\mu R$ is null is called an apparent horizon.

We now restrict attention to spherically symmetric CSS spacetimes.
The integral curves of the homothetic vector field (homothetic lines)
provide a natural fibration of the 4-dimensional spacetime, and
therefore a natural foliation of the reduced spacetime.  Labelling the
integral curves provides a natural global coordinate on the reduced
spacetime, which we shall call $z$. In this paper we construct
conformal diagrams for the reduced spacetime. Apart from any
singularities and conformal boundaries, lines of geometric
significance include the homothetic lines, the lines of constant $R$ and
the fluid world lines.

We shall describe two building blocks from which any spherically
symmetric CSS spacetime can be built. Our arguments in this section
are purely kinematic. They are therefore valid for any matter. On the
other hand, for a given type of matter not every spacetime that can be
assembled from the building blocks can be realized as a solution of
the field equations. We illustrate this in Section
\ref{section:causal} for perfect fluid matter.

The key kinematic consequence of self-similarity in general relativity
is the existence of a strong curvature singularity. It can be shown
from (\ref{CSS}) that in the direction $\xi^\mu$ any homothetic line
runs into a curvature singularity in finite proper time, proper
distance or affine parameter distance $s$, unless the curvature
vanishes identically on the curve \cite{LakeZannias,critcont}. If the
origin of $s$ is chosen so that the singularity is at $s=0$, the
Kretschmann scalar scales as $s^{-4}$, the Ricci scalar as $s^{-2}$
and the mass as $s^{-1}$. Every homothetic line is infinitely extended
in the direction $-\xi^\mu$ (i.e. with increasing $s$).

\begin{figure}
\includegraphics[width=6cm]{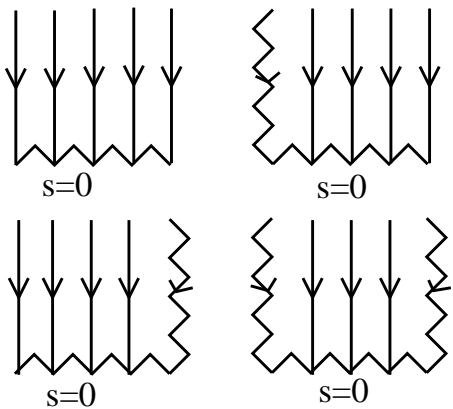}
\caption{The manifold structure of any spherically symmetric
self-similar spacetime: an interval $z_{\rm min}\le z\le z_{\rm max}$
times the half-line $0<s<\infty$. Note that this diagram is {\it not}
a conformal diagram: although $s=0$ is shown here as a horizontal line,
it could be a single point or a line with spacelike, timelike or null
segments.  Homothetic lines $z={\rm const.}$ are shown bold with
arrows that point to the singularity $s=0$. The four cases where none,
one or both of the limits $z=z_{\rm min}$ and $z_{\rm max}$ are
singularities are shown.}
\label{fig:manifold}
\end{figure}

Any homothetic spacetime has therefore the manifold structure of the
half-line $0<s<\infty$ crossed with an interval in $z$ (which can be
finite or infinite). $s=0$ is always a curvature singularity and
$s=\infty$ is always a physical infinity. The boundaries $z=z_{\rm min}$
and $z=z_{\rm max}$ can be either regular centers or central
singularities. The manifold structure is shown in
Fig.~\ref{fig:manifold}. It is important to note that this on its own
does not determine the causal structure, but it does mean that all
components of the singularity are connected. The usual conformal
diagram of a closed Friedmann universe with disjoint big bang and big
crunch singularities, for example, cannot be realized in a spherically
symmetric CSS spacetime. Nevertheless, fluid world lines can begin on
one section of the singularity (big bang) and end on another (black
hole/big crunch). The difference is that in a spherically symmetric
CSS solution these two components of the singularity must be connected
either directly or through a null singularity.

We obtain our two building blocks from considering radial null
geodesics that are also homothetic lines. Ori and Piran
\cite{OriPiran} call such a line a {\it simple} radial null geodesic
(SRNG). Assuming that the spacetime contains an isolated SRNG, with
neighboring homothetic lines are timelike on one side of the SRNG and
spacelike on the other, there are two possibilities: either the
homothetic lines converge with respect to radial null geodesics as
they approach the singularity at $s=0$ or they diverge.

\begin{figure}
\includegraphics[width=7.5cm]{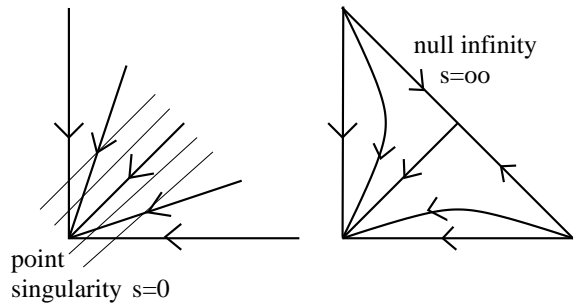}
\caption{A fan: a range of homothetic lines meet in a point on the
singularity. Homothetic lines are shown bold with arrows that point to
the singularity, in the direction $\xi^\mu$. One of them, $z=z_0$, is
also a radial null geodesic (a SRNG). Four other (non-simple) radial
null geodesics $z=z(s)$ are shown as thin lines. The conformal diagram
on the left is not compactified. On the right, the same conformal
diagram has been compactified to show that $(z=z_0,s=\infty)$ is a
piece of null infinity.}
\label{fig:fan}
\end{figure}

The first possibility is illustrated in Fig.~\ref{fig:fan}. We call
this building block a ``fan''. The homothetic lines converge as $s\to
0$.  The singularity $s=0$, for an interval of $z$, is then just a single
point.  Any non-simple radial null geodesic is given by $z=z(s)$. It
is clear from the figure that $z(s)$ is repelled from $z_0$ on either
side as $s\to 0$.  Conversely, $z(s)$ is attracted to $z_0$ as
$s\to\infty$.  $(z=z_0,s=\infty)$ is therefore not a point but an
extended piece of null infinity. In summary, the spacetime contains a
fan when it contains a SRNG $z=z_0$ such that $z(s)$ for non-simple
radial null geodesics approaches $z_0$ as $s\to\infty$, that is it
obeys $[z(s)-z_0]\ dz/ds<0$. A fan implies a point singularity at
$s=0$ for a finite range of $z$, and a null infinity at
$(s=\infty,z=z_0)$.

\begin{figure}
\includegraphics[width=7.5cm]{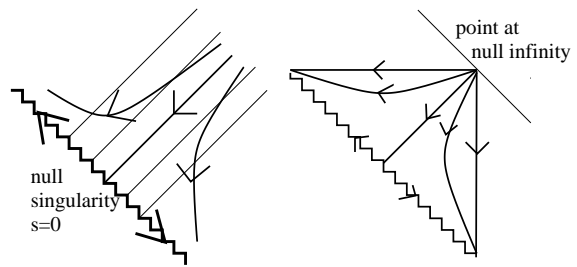}
\caption{A splash: A SRNG $z=z_0$ is flanked by homothetic lines that
diverge from it as the singularity is approached. Homothetic lines
$z={\rm const.}$ are shown bold with arrows that point to the
singularity. The SNRG branches and runs along the singularity. The
singularity $s=0$ is null and is at finite distance.  The conformal
diagram on the left is not compactified. The conformal diagram on the
right is compactified to show that a range of homothetic lines meet in
a point at null infinity.}
\label{fig:splash}
\end{figure}

The opposite structure is illustrated in Fig.~\ref{fig:splash}. Here
the non-simple radial null geodesics $z=z(s)$ approach $z=z_0$ as
$s\to 0$, and are repelled from $z_0$ as $s\to\infty$. Therefore the
singularity $(z=z_0,s=0)$ is reached not just by one but many radial
null geodesics, and so it is a null line. Conversely, $s=\infty$ for a range of $z$ on each side of $z_0$
is a just a single point, rather than an extended piece, of null
infinity. We call this structure a ``splash'' because the
homothetic line $z=z_0$ runs into the singularity at $s=0$ and then
bifurcates to form the null singularity. The spacetime contains a splash
when it contains a SRNG $z=z_0$ such that $z(s)$ for non-simple radial
null geodesics approaches $z_0$ as $s\to 0$. A splash implies a null central
singularity at $s=0$.

Note that the coordinates $s$ and $z$ break down at both $s=0$ and
$s=\infty$ in a fan or splash. Regular double null coordinates for a
fan and a splash are constructed in \cite{critcont}. (This
construction also proves that the singularity $s=0$ in a splash is
null, a fact which was asserted but not proved in \cite{OriPiran} and
\cite{Nolan}.)  Note also that in principle the homothetic lines could
be timelike (or spacelike) on {\it both} sides of the SNRG. In this
case, one would only have the upper (or lower) half of the diagrams
shown in Fig.~\ref{fig:fan} and Fig.~\ref{fig:splash}, corresponding
to what one might term a ``semi-fan'' or ``semi-splash''
respectively. However, we do not show the corresponding figures
explicitly.


\section{Perfect fluid matter}
\label{section:dynamics}


We now examine spherically symmetric CSS spacetimes with
perfect fluid matter with the equation of state $p=\alpha\mu$, where
$p$ is the pressure, $\mu$ is the total energy density, and
$0\le\alpha<1$ is a constant. We briefly introduce notation based on
\cite{CarrColey} and \cite{BicknellHenriksen}. The metric is given in
comoving coordinates by
\begin{equation}
ds^2=-e^{2\nu}dt^2+e^{2\lambda}dr^2+R^2\,d\Omega^2,
\end{equation}
where $0\le r<\infty$ labels fluid worldlines. The fluid velocity is
therefore $U^\mu=(e^{-\nu},0,0,0)$. Spherically symmetric CSS
solutions can be put into a form in which all dimensionless quantities
are functions only of the dimensionless self-similar variable $z=r/t$.
In particular, we enforce homotheticity by demanding that
\begin{eqnarray}
\nonumber
&\nu(t,r)=\nu(z), \qquad
&\lambda(t,r)=\lambda(z), \\
&R(t,r)=rS(z), \qquad 
&8\pi G\mu(t,r) =r^{-2}\eta(z).
\end{eqnarray}
The homothetic vector defined by (\ref{CSS}) is
\begin{equation}
\xi^\mu{\partial \over \partial ^\mu}=-t{\partial\over\partial t}
-r{\partial\over \partial r}
\end{equation}
in these coordinates. Varying $z$ for a given $t$ specifies the
spatial profile of various quantities. For a given value of $r$ (i.e.,
for a given fluid element), it specifies their time evolution. The
remaining gauge-freedom in CSS solutions corresponds to a scaling
of the $r$ and $t$ coordinates.

Two quantities have special physical significance
\cite{CarrColey}. The first is the CSS mass function, which is defined
in terms of the Hawking mass $m(r,t)$ and the area radius $R(t)$:
\begin{equation}
M(z)={m\over R}.
\end{equation}
One has an apparent horizon wherever $M=1/2$.
The second is the CSS velocity function
\begin{equation}
V(z)=e^{\lambda -\nu}z ,
\end{equation}
which represents the velocity of the spheres of constant $z$ relative to the
fluid. In geometric terms,
\begin{equation}
{\xi^\mu u_\mu \over \sqrt{\xi^\nu \xi_\nu}}={1\over \sqrt{1-V^2}}.
\end{equation}
This should not be confused with the velocity of the fluid with
respect to the lines of constant $R$, the ``radial 3-velocity'', which
we denote by $V_R$. Special significance is attached to values of $z$
for which $|V|=\sqrt{\alpha}$ and $|V|=1$. The first corresponds to a
sonic point, the second to a black-hole event horizon or a
cosmological particle horizon. A useful relation is that for radial
null geodesics
\begin{equation}
\label{RNG}
d\ln r={d\ln z\over 1\pm V(z)},
\end{equation}
where the + and - signs correspond to ingoing and outgoing geodesics, respectively. This again shows that lines $z=z_0$ where $V(z_0)=\pm 1$ are both
radial null geodesics and homothetic lines, that is they are SRNGs.

It should be noted that finite values of $z$ sometimes correspond to
zero or infinite $R$ and zero values of $z$ sometimes correspond to
non-zero $R$. Also some solutions span both negative and positive
values of $t$, which means that $z=r/t$ jumps from $-\infty$ to
$+\infty$. (This occurs at $t=0$, which is the homothetic line
orthogonal to the fluid worldlines.)

Two of the Einstein equations can be solved in closed form to give $\nu$
and $\lambda$ in terms of $z$, $S(z)$ and $\eta(z)$. One then obtains
a second-order ODE for $S$ and a first-order ODE for $\eta$ in the
independent variable $z$. The equation for $d^2S/dz^2$ can be written
as two first-order equations for $dS/dz$ and $dM/dz$. One then obtains
the system 
\begin{eqnarray}
{dM\over dz}&=&a(S,M,\eta,z), \\
{dS\over dz}&=&b(S,M,\eta,z), \\
{d\eta\over dz}&=&{g(S,M,\eta,z)\over V^2(S,\eta,z)-\alpha}, 
\end{eqnarray}
subject to the constraint
\begin{equation}
c(S,M,\eta,z)=0.
\end{equation}
Here $a$, $b$, $c$, $V^2$ and $g$ are known regular functions of $M$,
$S$, $\eta$ and the independent variable $z$.  The constraint can in
principle be used to eliminate $z$ from the right-hand sides, so that
one obtains a 3-dimensional dynamical system. Therefore, for a given
value of $\alpha$, the space of solutions is locally 2-dimensional. In
practice the constrained and non-autonomous form is retained for
calculations.

Points where $V=\pm\sqrt{\alpha}$ need particular attention. In phase
space the condition $V^2=\alpha$ corresponds to a 2-dimensional
surface, which we call the {\it sonic surface}. Eqn (10) implies that
solutions can be continued through the sonic surface only on the
1-dimensional line in phase space where both $V^2=\alpha$ and $g=0$;
we call this the {\it sonic line}. We use the term {\it sonic point}
for the 3-dimensional surface in spacetime where $V^2=\alpha$.

The details are summarized for completeness in Appendix
\ref{appendix:sonic}. Here we discuss the effect that the existence of
sonic points has on the local dimensionality of the solution
space. The sonic line can be divided into three parts. It has a
``focal'' segment that cannot be crossed by any solution curves and is
a repeller. It also has a ``saddle segment'', each point of which is
crossed by two isolated analytic solutions. Finally, it has two
``node'' segments, each point of which is crossed by a 1-parameter
family of $C^1$ solutions and an isolated solution. All the $C^1$
solutions are tangential to each other at the sonic line. Therefore,
if one does not require the solution to be analytic, any of the $C^1$
solutions on one side of the sonic surface can be continued by any of
the $C^1$ solutions on the other side.  Like a saddle, a node is also
crossed by two analytic solutions. One of these is a member of the $C^1$
family, while the other is the isolated solution.

The global dimensionality of the families of solutions that we discuss
below can be determined as follows. If $z_{\rm min}$ is a singularity,
the space of solutions locally has two free parameters. If it is a
regular center, it obeys one condition and so the solution locally has
only one free parameter. If the solution crosses the sonic point
analytically (i.e. at a node or a saddle), it loses one parameter. For
example, critical solutions have a regular center and are analytic at
the sonic point, so there are no free parameters, indicating that
there are at most isolated solutions with those properties. On the
other hand, if the solution crosses a node as a $C^1$ solution, the
continuation is not unique and so gains one parameter. The same
argument applies to any subsequent sonic point.

Note that the presence of a sonic point cannot be inferred from the
conformal diagrams presented below. Each sonic point is a timelike
homothetic line, but a region in which the homothetic lines are
timelike need not contain a sonic point.


\section{Causal structure of perfect fluid CSS solutions}
\label{section:causal}


Carr and Coley \cite{CarrColey} (from now on CC) classify solutions in
terms of their behavior in the asymptotic limit
$|z|\rightarrow\infty$, since this can only take one of a few simple
forms: asymptotically Friedmann, asymptotically Kantowski-Sachs, what
they term asymptotically `quasi-static', and asymptotically
Minkowski. There is also a family which asymptotes to the Minkowski
form at finite $z$ because $R\rightarrow\infty$ there.  In the
conformal diagram, all these asymptotic behaviors are associated with
the region near $i^o$.  CC also discuss the behavior of solutions in
the limit $z\rightarrow0$ and find that they are either exactly static
or asymptotically Friedmann (corresponding to a `regular'
origin). There are also solutions which have their origin at finite
$z$ because $R\rightarrow 0$ there, corresponding to a spacelike
singularity. In the conformal diagram, these asymptotic behaviors are
associated with the regions near $i^+$ or $i^-$.

CC place considerable emphasis on the form of the scale factor $S(z)$
since this specifies the physical behavior of the fluid elements. They
also emphasize the form of the velocity function $V(z)$ since this
identifies event and particle horizons, sonic points and physical
singularities. Although the $V(z)$ representation does not yield a
complete understanding of solutions -- one is projecting solutions in
a 3-dimensional space onto a particular 2-dimensional plane, so that
many physically distinct solutions may be superposed -- it uniquely
specifies features relevant to the conformal structure. In particular,
$|V|=1$ corresponds to a null line and infinite $|V|$ corresponds to a
singularity. Integrating Eq. (8)  shows that the value of $r$
at $V=1$ generally diverges for any outgoing photon, so this corresponds to $I^+$.
Our strategy here is to present the form of the functions
$S(z)$ and $V(z)$ for each of the types of solution and then to use
these to construct the conformal diagram. We will also show some fluid
lines and homothetic lines in each case.

Note that we can trivially set $t\to -t$ and hence $z\to -z$ in any
CSS solution, and so obtain its time reverse. As a matter of
convention, we shall present each solution in only one time
orientation, with ``big bang'' singularities in the past and ``black
hole'' singularities in the future. For example, Fig. \ref{fig:af1}
describes an open universe that starts with a big bang and expands
forever. Turned upside down, the same solution describes gravitational
collapse from a regular initial state \cite{OriPiran}.

Carr et al. \cite{CCGNU1,CCGNU2} (from now on CCGNU) classify
solutions in terms of the global causal properties of the homothetic
vector $\xi^\mu$ and this is particularly relevant in
constructing the conformal diagram. The solutions are found to have
five possible forms. There are type-T solutions for which $\xi^\mu$ is
always timelike; these develop a shock, since their orbits necessarily
end at an irregular sonic point. There are type-S solutions for which
$\xi^\mu$ always is spacelike; these always have $|V|>1$ and include
some asymptotically Friedmann and asymptotically quasi-static
solutions. There are type-TS solutions, such as the flat Friedmann
one, for which $\xi^\mu$ changes causality once. Finally there are are
type-TST, STST, and STSTS solutions with two, three and four causality
changes, respectively.  

Here we make this shorthand description more detailed by denoting the
fans and splashes that separate the S and T regions by {f} and
{p}. Finally we denote the two boundaries $z_{\rm min}$ and $z_{\rm
max}$ by either r for a regular center or s for a singularity. Note
that the upper case letters S and T denote regions, while the lower
case letters f, p, r and s denote lines. All kinematically possible
CSS spacetimes can be constructed systematically from an alphabet of
six letters subject to the following rules: The first and last letter
are r or s. The other letters are f, p, T, S. Capital letters and
lower case letters must alternate.


\subsection{Asymptotically Friedmann solutions}\label{sec:af} 

\begin{figure}
\includegraphics[width=6.5cm]{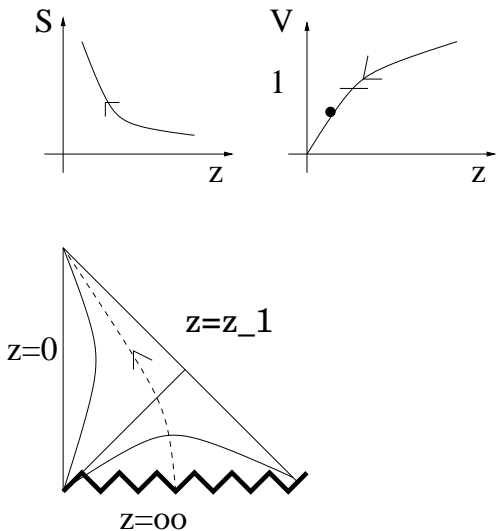}
\caption{The form of $S(z)$ and $V(z)$ for the asymptotically
Friedmann solutions which expand forever, together with the conformal
diagram. Homothetic lines are shown, together with a single fluid world line
(shown broken). The arrows, here and in all diagrams that
follow, indicate the direction of time. The causal structure is
``singularity, region where the homothetic lines are spacelike, fan,
timelike region, regular center'', or sS{f}Tr in our shorthand
notation.}
\label{fig:af1} 
\end{figure}

There are two 1-parameter families of asymptotically Friedmann
solutions as $|z|\rightarrow\infty$, one with $z>0$ and the other with
$z<0$. The parameter characterizing these solutions measures the
underdensity or overdensity relative to the flat Friedmann solution
and is also associated with the asymptotic energy $E$. The $z>0$
solutions correspond to inhomogeneous models that start from an
initial big bang singularity at $z=\infty$ ($t=0$) and then, as $z$
decreases, either expand to infinity or recollapse. The $z<0$
solutions are just the time-reverse of these but we do not discuss
them explicitly. Depending on the value of $E$, two qualitatively
different types of solution can be distinguished:

1) All solutions which are underdense ($E>0$) or insufficiently
overdense ($E<0$ but exceeding some negative critical value $E_{\rm
crit}$) reach the sonic surface at $V=\sqrt{\alpha}$. They have $V=1$ at 
some point $z_1$ and this corresponds to the cosmological particle horizon. The solutions in
this class are of type ST for $z>0$ and are illustrated in Fig.
\ref{fig:af1}. Those which reach the sonic line in the range of $z$
corresponding to nodes may
be attached to a subsonic ($V<\sqrt{\alpha}$) solution which is
regular solution at the origin. These solutions are also described by
a single parameter and this is a measure of the density at the origin
$z=0$, so the transonic solutions represent density fluctuations that
grow at the same rate as the particle horizon. They are generally
non-analytic at the sonic point. While there is a continuum of regular
underdense solutions, regular overdense solutions only occur in narrow
bands (with just one solution per band being analytic). The overdense
solutions exhibit oscillations in the subsonic region and solutions
with larger numbers of oscillations form ever narrower bands within the
one-oscillation band (in terms of the ranges of $z$ at the sonic
line). The higher bands are all nearly static near the sonic point,
although they depart from the static solution as they approach the
origin. The conformal diagram for these solutions is essentially
identical to that for flat Friedmann itself.


\begin{figure}
\includegraphics[width=6.5cm]{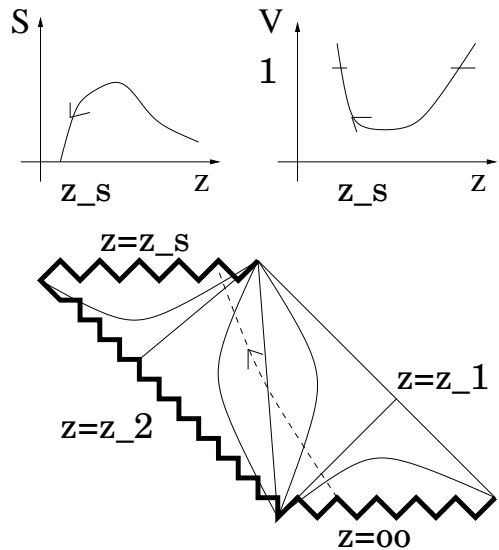}
\caption{The form of $S(z)$ and $V(z)$ for the asymptotically
Friedmann solutions which contain a black hole with an event horizon,
together with the conformal diagram. There is a fan at $z=z_{\rm ph}$
and a splash at $z=z_{\rm eh}$, and the causal structure is
sS{f}T{p}Ss.}
\label{fig:af2}
\end{figure}

\begin{figure}
\includegraphics[width=6.5cm]{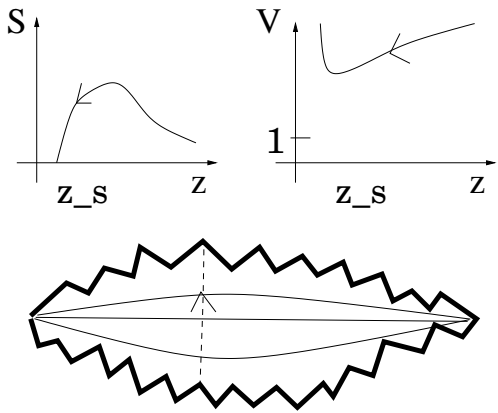}
\caption{The form of $S(z)$ and $V(z)$ for the asymptotically
Friedmann solutions which contain a ``universal'' black hole, together
with the conformal diagram. The causal structure is sSs, with no fan
or splash, and homothetic lines are spacelike everywhere.}
\label{fig:af3} 
\end{figure}

2) Solutions that are sufficiently overdense with respect to the flat
Friedmann solution (i.e. for $E$ less than the critical value $E_{\rm
crit}$) have the feature that $V$ reaches a minimum and then rises
again to infinity as $z$ decreases to some finite value $z_S$. This
indicates the formation of a singularity at which $S\rightarrow0$ and
$\mu\rightarrow\infty$ for finite a value of $z$. The solutions in
this class are of type STS or S and are illustrated in
Fig.~\ref{fig:af2} and Fig.~\ref{fig:af3}, respectively. They
correspond to black holes growing at the same rate as the
Universe. [Note that the type STS solution with the smallest value of
$z_S$ is the one for which the minimum value of $V$ is
$1/\sqrt{\alpha}$ and this solution must touch the sonic surface at
the value of $z$ associated with the saddle/node transition.]
Providing $E$ exceeds some other critical value $E_*$, the minimum of
$V$ is below 1, as illustrated in Fig.~\ref{fig:af2}, so there is a
black-hole event horizon at $z_2$ and a cosmological particle horizon at $z_1$ where
$V=1$. However, it should be stressed that the conformal diagram is
very different from that for a non-self-similar black hole in a flat
Friedmann background.  In particular, the black hole singularity is
connected to the big bang singularity and necessarily naked for a
while. For $E<E_*$, the minimum of $V$ is above 1, as illustrated in
Fig.~\ref{fig:af3}, and the entire Universe is inside the black
hole. (However, there is always an apparent horizon since CC show that
the minimum of $M$ is necessarily below $1/2$.) In this case, the
conformal diagram has two spacelike singularities and there are no
null infinities. This is because the integral over (\ref{RNG}), for
both signs, converges as $z\to\infty$ and $z\to z_S$, so any null
geodesic covers only a finite range of $\ln r$ between the big bang
singularity and the black hole/big crunch singularity. This contrasts
with the conformal diagram for a recollapsing Friedmann universe,
which has two separated spacelike singularities. 

Note that when the minimum of $V$ exactly equals 1, the conformal
diagram in Fig.~\ref{fig:af2} loses the part in which the homothetic
lines are timelike and so there is no longer a naked singularity. In
terms of causal structure, the ``fTp'' part of Fig.~\ref{fig:af2} is
replaced by a ``semi-fan'' and a ``semi-splash''. However, the
conformal diagram is still very different from the one shown in
Fig.~\ref{fig:af3}, so there is a qualitative change as the minimum of
$V$ passes through 1. In the cases discussed below, similar
considerations apply whenever $V$ has an extremum at $\pm 1$, but we
will not comment on this explicitly.


\subsection{Asymptotically quasi-static solutions}


As discussed by CC, there is exactly one self-similar static solution
for each value of $\alpha$ and a 1-parameter family of solutions that
are asymptotically static (in the sense that the radial 3-velocity
$V_R$ tends to zero as $|z|\rightarrow\infty$). There is also a
2-parameter family of solutions that are asymptotically ``quasi-static''
(in the sense that $V_R$ is finite but not necessarily zero). However,
it should be emphasized that the solutions in this class may only be
close to the static solution for part of their evolution. One of the
two parameters can be taken to be the asymptotic energy $E$, while the
other (denoted by $D$) gives the value of $z$ at the big bang
or big crunch singularity. Note that $D$, unlike $E$ and $V_R$, cannot
be expressed explicitly in terms of the asymptotic parameters, so it
is sometimes more convenient to use $V_R$ as the second parameter.

The key feature of these asymptotically quasi-static solutions is that
they span both negative and positive values of $z$ and necessarily
pass from $z=-\infty$ to $z=+\infty$, whereas the asymptotically
Friedmann solutions are confined to $z>0$ or $z<0$ and are
symmetric. This is because the big bang occurs at $z=-1/D$
(corresponding to a negative value of $t$) in these solutions, so the
limit $|z|\rightarrow\infty$ has no particular physical
significance. These solutions can be interpreted as inhomogeneous
cosmological models with an advanced big bang. Equivalently, for the
time-reversed solutions, there is a big crunch singularity at
$z=+1/D$. There are two types of asymptotically quasi-static
solutions.  Those with $E$ exceeding some negative critical value
$E_{\rm crit}(D)$ expand or collapse forever. Those with $E<E_{\rm
crit}(D)$ expand and then recollapse.

\begin{figure}
\includegraphics[width=6.5cm]{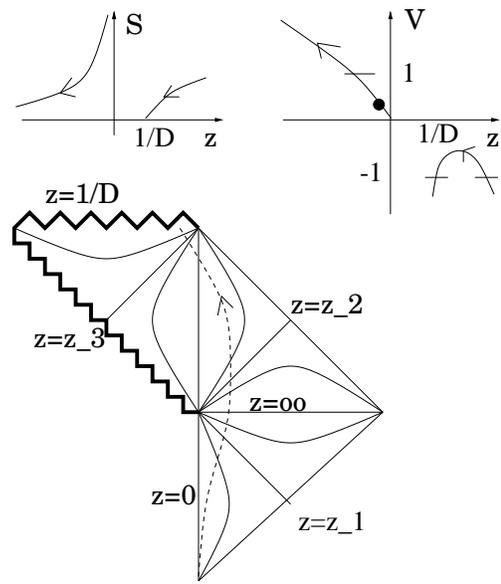}
\caption{The form of $S(z)$ and $V(z)$ for the ever-collapsing
asymptotically quasi-static solutions which contain a naked
singularity, together with the conformal diagram. The causal structure is
rT{f}S{f}T{p}Ss.}
\label{fig:aqs1}
\end{figure}

\begin{figure}
\includegraphics[width=6.5cm]{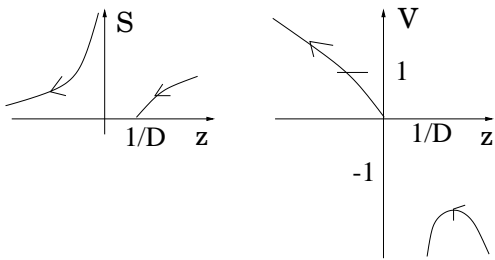}
\caption{The form of $S(z)$ and $V(z)$ for the ever-collapsing
asymptotically quasi-static solutions which do not contain a naked
singularity. The conformal diagram is the same as in
Fig.~\ref{fig:af1}, with structure sS{f}Tr, although upside down.}
\label{fig:aqs2}
\end{figure}

1) Ever-collapsing solutions. These start out from an infinitely
dispersed state and describe the collapse of an inhomogeneous gas
cloud to a singularity at $z=+1/D$ (i.e., after $t=0$). (The
ever-expanding solutions are just the time reverse of these and will
not be considered explicitly.) They are of type TS or TSTS and
illustrated in Fig.~\ref{fig:aqs1} and Fig.~\ref{fig:aqs2}.  They
start with $V=0$ at $z=0$ and then, as $z$ decreases, reach a sonic
point where $V=\sqrt{\alpha}$. In this context, it should be noted
that the second parameter $D$ has relatively little effect on the form
of the solutions in the subsonic regime. Indeed, all solutions apart
from the exactly static one must be asymptotic to the flat Friedmann
solution at small $|z|$. In particular, the models can collapse from
infinity (i.e. $S\rightarrow\infty$ as $z\rightarrow0$ or
$t\rightarrow -\infty$) only if $E$ is positive or lies in discrete
bands if negative. The subsonic solution is then attached to a
supersonic asymptotically quasi-static solution at the sonic line. The
supersonic solutions pass through a Cauchy horizon at $z_1$ (where $V=1$)
before tending to the quasi-static form at $z=-\infty$ and jumping to
$z=+\infty$. As $z$ further decreases, $V$ first reaches a maximum and
then diverges to minus infinity when it encounters the singularity at
$z=1/D$. The maximum will be above -1 if $E$ is less than some negative
value $E_+(D)$ and there will then be two points $z_2$ and $z_3$ where $V=-1$. This case is illustrated in Fig.~\ref{fig:aqs1} and
the conformal diagram is a combination of the one shown in
Fig.~\ref{fig:af1} and Fig.~\ref{fig:af2}. This shows that one
necessarily has a naked singularity, as pointed out by Ori and Piran.
The maximum of $V$ will be below -1 for $E>E_+(D)$ and this case is
illustrated in Fig.~\ref{fig:aqs2}. The conformal diagram now
resembles the one in Fig.~\ref{fig:af1}, except that the singularity
has $z=1/D$ rather than an infinite value of $z$.


\begin{figure}
\includegraphics[width=6.5cm]{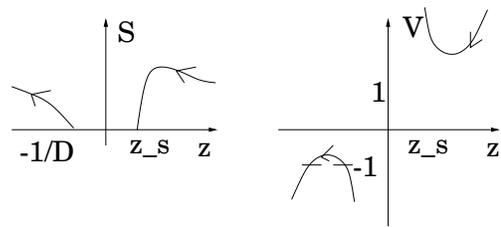}
\caption{The form of $S(z)$ and $V(z)$ for the recollapsing
asymptotically quasi-static solutions which contain two singularities,
one of which is naked. This shows the case $V_{\rm max}>-1$ and
$V_{\rm min}>1$. Both this case and the case with $V_{\rm max}<-1$ and
$V_{\rm min}<1$ have the conformal diagram of Fig.~\ref{fig:af2} except that the
big bang singularity is at $z=-1/D$ instead of $z=\infty$.}
\label{fig:aqs3}
\end{figure}

\begin{figure}
\includegraphics[width=6.5cm]{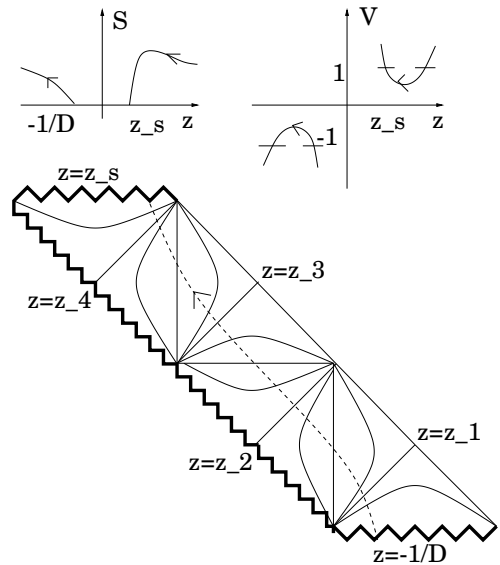}
\caption{The form of $S(z)$ and $V(z)$ for the recollapsing
asymptotically quasi-static solutions which contain two naked
singularities, together with the conformal diagram. The structure is
sS{f}T{p}S{f}T{p}Ss.}
\label{fig:aqs4}
\end{figure}

\begin{figure}
\includegraphics[width=6.5cm]{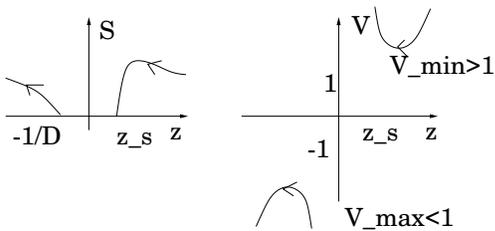}
\caption{The form of $S(z)$ and $V(z)$ for the recollapsing
asymptotically quasi-static solutions which contain no naked
singularities. The conformal diagram is the same as in
Fig.~\ref{fig:af3}, with structure sSs.}
\label{fig:aqs5}
\end{figure}

2) Expanding and recollapsing solutions. Solutions with $E$ less than
the critical value $E_{\rm crit}(D)$ expand from an initial
singularity in the $z<0$ region and then recollapse to another
singularity in the $z>0$ region.  They are of type S, STS or
STSTS. The value of $z$ is $-1/D$ at the initial singularity but
depends on both $E$ and $D$ at the final one.  As $z$ decreases from
$-1/D$, $V$ rises from $-\infty$, reaches a maximum below
$-\sqrt{\alpha}$ and then tends to the quasi-static form as
$z\rightarrow -\infty$. The solution then jumps to $z=+\infty$ and
enters the $z>0$ regime. As $z$ continues to decrease, $V$ decreases
to a minimum above $\sqrt{\alpha}$ and then tends to $+\infty$ at the
value of $z$ corresponding to the recollapse singularity. We now have
four possible situations, depending on whether the minimum of $V$ is
above or below $+1$ and whether the maximum of $V$ is above or below
$-1$. If $V_{\rm max}>-1$ and $V_{\rm min}>1$, one necessarily has a
black-hole event horizon at $z_1$ and a cosmological particle horizon at $z_2$ in the
$z<0$ regime. This occurs if $E$ exceeds some negative value $E_*(D)$
and the minimum of $V$ will reach $\sqrt{\alpha}$ when $E$ reaches
$E_{\rm crit}(D)$, corresponding to the last recollapsing solution. In
this case, the solution is of the STS type and illustrated in
Fig.~\ref{fig:aqs3}. The conformal diagram just resembles that in
Fig.~\ref{fig:af2}, except that the initial singularity is at $z=1/D$ 
rather than $z=\infty$, and this shows that the black hole singularity is
necessarily naked. The case with $V_{\rm max}<-1$ and $V_{\rm min}<1$
is equivalent to this and not shown explicitly. The case with $V_{\rm
max}>-1$ and $V_{\rm min}<1$ is of the STSTS type and illustrated in
Fig.~\ref{fig:aqs4}. There are now four points ($z_1,z_2,z_3,z_4)$ at which $V=1$
and so the conformal diagram is more complicated, though the
singularity structure is unaltered. The case with $V_{\rm max}<-1$ and
$V_{\rm min}>1$ is illustrated in Fig.~\ref{fig:aqs5}. There are now
no points where $V=1$ and so the conformal diagram is as in
Fig.~\ref{fig:af3}.  In principle, there might be solutions with
$E<E_{\rm crit}(D)$ in which $V$ has neither a maximum nor a
minimum. However, such solutions would have two sonic points and
parameter-counting suggests that it is unlikely that any solution
having crossed the sonic surface once in a saddle could cross it
again.


\begin{figure}
\includegraphics[width=7.5cm]{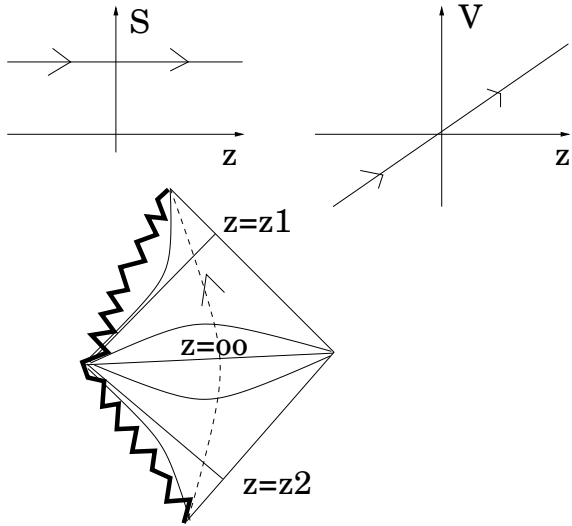}
\caption{The form of $S(z)$ and $V(z)$ for the exact static 
solution, together with the conformal diagram. The structure is
sT{f}S{f}Ts.} 
\label{fig:aqs6} 
\end{figure}

The exact static solution is illustrated in Fig.~\ref{fig:aqs6}. The
conformal diagram has the Minkowski form at large distance but there is a
timelike naked singularity at the origin and points $z_1$ where $V=-1$ and $z_2$ where $V=1$. Note that there are no
solutions which are asymptotic to the static solution at $z=0$ apart
from the static solution itself, so there are no perturbations of the
exact static solution. The asymptotically quasi-static solutions with
a regular centre are asymptotically Friedmann at $z=0$.


\subsection{The asymptotically Minkowski solutions with $\alpha>\frac{1}{5}$}


There are two families of asymptotically Minkowski solutions in the
$z>0$ regime and both are physical only for
$\alpha>\frac{1}{5}$. Equivalent (time reversed) solutions exist in
the $z<0$ regime but we do not discuss these explicitly.

\begin{figure}
\includegraphics[width=7.5cm]{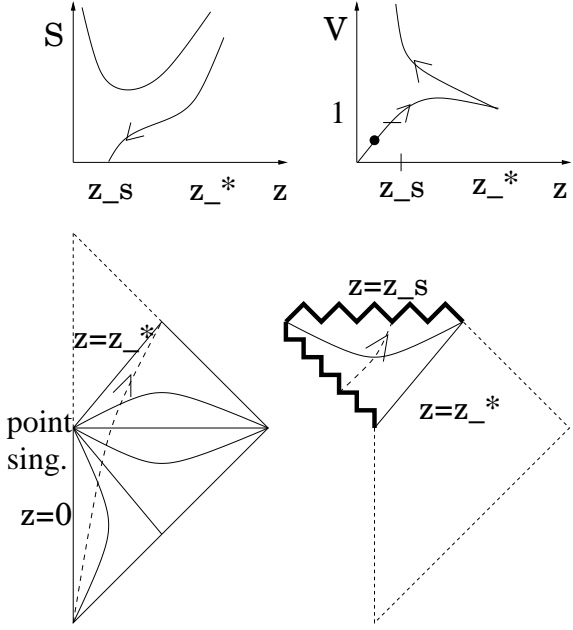}
\caption{The form of $S(z)$ and $V(z)$ for the 2-parameter bouncing
and collapsing asymptotically Minkowski solutions, together with the
conformal diagram. There are two solutions which are geodesically
incomplete. Each of them can separately be continued as Minkowski, and
this is shown by the thin dashed line. One fluid world line is shown as
a thicker dashed line.  The two solutions can also be glued together
at $z=z_s$. Note that in this case $z=z_*$ is a fan on one side and a
splash on the other, because $V(z)$ has a minimum at $V=1$ there,
rather than crossing $V=1$. The causal structure of the incomplete
spacetime on the left is rT{f}S{f}T{f}. The causal
structure of the incomplete spacetime on the right is {p}Ss.}
\label{fig:am1}
\end{figure}

A) The first family is described by two parameters and has
$V\rightarrow1$, $S\rightarrow\infty$ and $\mu\rightarrow0$ at some
{\it finite} value $z=z_*$. In this limit,
\begin{equation}
\dot{V}/V=-(5\alpha-1)/(1-\alpha) .
\end{equation}
This family is illustrated in Fig.~\ref{fig:am1}. Although the value
of $z$ is finite at $z_*$, the Schwarzschild radial distance ($R=rS$)
is infinite, as is clearly seen in the $S(z)$ diagram. These are the
``explosive solutions'' of Ori and Piran \cite{OriPiran}, in which the
fluid velocity approaches the speed of light, leaving null data for
flat spacetime on an outgoing null cone.

\begin{figure}
\includegraphics[width=7.5cm]{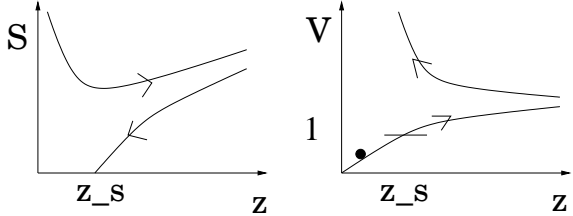}
\caption{The form of $S(z)$ and $V(z)$ for the 1-parameter bouncing
and collapsing asymptotically Minkowski solutions. The conformal
diagram and the causal structure are the same as in
Fig.~\ref{fig:am1}, with $z_*$ replaced by $z=\infty$.}
\label{fig:am2}
\end{figure}

B) The second family is described by one parameter and has
$V\rightarrow V_*>1$, $S\rightarrow\infty$ and $\mu\rightarrow0$ as
$z\rightarrow\infty$.  The expression for $V_*$ is
\begin{equation}
\label{V*}
V_*=\frac{\alpha(\alpha+1)+\sqrt{\alpha(\alpha^3-\alpha^2+3\alpha+1)}}
{1-\alpha} .
\end{equation}
This family is illustrated in Fig.~\ref{fig:am2}. Note that the $V_*$
decreases as $\alpha$ decreases and reaches $1$ when $\alpha=1/5$.

In both cases, the limit $S\rightarrow\infty$ can be regarded as
corresponding to an infinitely dispersed state, analogous to the late
stage of an open Friedmann model. Also both families of solutions have
$m/R\rightarrow0$ in the asymptotic limit. The mass $m$ itself tends
to a finite value in case A but to zero in the case B. It
is important to realize that these solutions are not asymptotically
flat in the usual sense. Rather, they are perfect-fluid spacetimes for
which the Minkowski geometry is obtained asymptotically along certain
coordinate lines. This is reminiscent of the open Friedmann solution,
which asymptotically approaches the Milne model along certain
time lines.

As can be seen in Figs.~\ref{fig:am1} and \ref{fig:am2}, both families
contain solutions that can be connected either to the origin $z=0$
(via a sonic point) or to a singularity (for which $S\rightarrow0$ at
some finite value $z_S$). The former are of type T or ST and correspond to
singularity-free models which collapse and then bounce at some finite
density into another expansion phase. The latter are of type S and
collapse to a singularity without passing through a sonic point. We
therefore have four situations. The bouncing 2-parameter case is
illustrated in the left conformal diagram of Fig.~\ref{fig:am1}.  The
conformal structure is very similar to the Minkowski case and can
indeed be extended to a solution whose conformal structure is exactly
like Minkowski if one adds another Minkowski patch to the top-left
corner (shown dotted).  There is a point singularity at $r=t=0$. Since
$\mu r^2=0$ at $z=z_*$, there is a singularity only where
$r=0$.  The conformal diagram for the collapsing 2-parameter case is shown on the right of Fig.~\ref{fig:am1}. This could also be attached to a Minkowski region
(shown dotted), in which case it is similar to the conformal diagram in
Fig.~\ref{fig:aqs1}.  It might also in principle be patched on to the top corner of the left conformal diagram. The 1-parameter case is
illustrated in Fig.~\ref{fig:am2}. Although $V=V_*>1$ is finite as
$z\to\infty$, the surface $z=\infty$ is not actually spacelike but
consists of two null branches. As shown in appendix
\ref{appendix:MinkowskiB}, one of them is an outgoing null surface at
finite distance -- as in the type A solutions, it is possible to glue
it onto Minkowski here -- and the other branch is $I_+$.


Note that there is a 1-parameter family of singular solutions for each
value of $z_S$. Most of these will be asymptotic to either a type-A
asymptotically Minkowski solution or a quasi-static solution. However,
for sufficiently large values of $z_S$, there will also be one type-B
asymptotically Minkowski solution and one asymptotically Friedmann
solution. The limiting value of $z_S$ is the same in each case,
reflecting the fact that the conditions at large values of $z$ have
little influence on what happens near the singularity.


\subsection{Asymptotic Kantowski-Sachs solutions}


The final class of models is associated with the Kantowski-Sachs
solution. For each $\alpha$ there is a unique self-similar
Kantowski-Sachs solution and there also exists a $1$-parameter family
of solutions asymptotic to this at both large and small values of
$|z|$. Solutions with $-1/3 < \alpha <1$ are probably unphysical
because the mass is negative and they are also tachyonic for $0 <
\alpha <1$.  Solutions with $-1< \alpha < -1/3$ avoid these
unsatisfactory features. Although such equations of state violate the
strong energy condition, they could could well arise in the early
Universe due to inflation or particle production effects. 

\begin{figure}
\includegraphics[width=6cm]{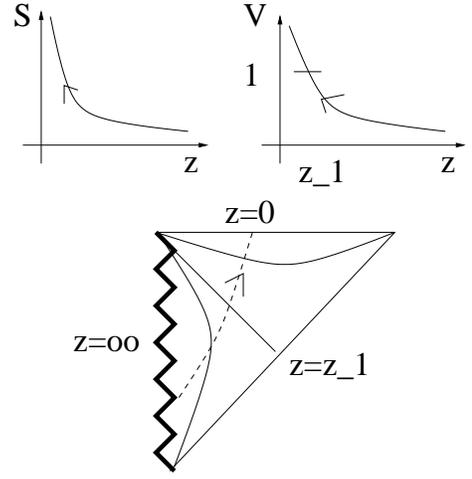}
\caption{The form of $S(z)$ and $V(z)$ for the asymptotically
Kantowski-Sachs solutions with $-1< \alpha < -1/3$, the only physical ones. 
The causal structure is sT{f}Sr.}
\label{fig:ks}
\end{figure}

In the asymptotically KS solutions, $S\propto z^{-1}$ and $V$ is also
a decreasing power of $z$ for $-1< \alpha < -1/3$. Thus both $S$ and
$V$ decrease to $0$ as $z\to \infty$, as indicated in
Fig.~\ref{fig:ks}. This means that there is a timelike singularity at
$t=0$ and there is no spacelike infinity since $R$ depends only on
$t$. The conformal diagram is therefore as indicated in Fig.~\ref{fig:ks}.  The
KS solution can be obtained from the static one by changing the value
of $\alpha$ and interchanging $r$ and $t$; this explains why the
conformal diagram of the KS solution resembles one half of that of the
static solution \cite{CarrColey}.


\section{Critical solutions and shocks}
\label{section:critical}


Critical phenomena in gravitational collapse arise during the
evolution of asymptotically flat regular initial data that are close
to the threshold between black hole formation and dispersion.  Such
near-critical data approach a universal intermediate attractor before
they either form a black hole or disperse. In dynamical systems terms,
this ``critical solution'' is an attractor of codimension one, whose
stable manifold separates the phase space into solutions that form a
black hole and those that disperse. In terms of perturbation
theory, it must have exactly one unstable linear perturbation mode,
such that adding any amount of the unstable mode with one sign
eventually drives the solution to collapse, and adding any amount with
the opposite sign drives it to dispersion.

The property of being at the boundary of collapse can be realized in
two ways: either the solution is static, corresponding to an unstable
star, or it is self-similar, corresponding to scale-invariant
collapse to a point. In ``type II'' critical phenomena the
critical solution is self-similar (either CSS or DSS), and the black
hole mass $m_{\rm bh}$ as a function of the initial data exhibits a
power-law scaling of the form
\begin{equation}
m_{\rm bh}\sim (p-p_*)^\gamma.
\end{equation}
Here $p$ is any one parameter of the initial data that is being varied
while all other parameters remain fixed, and $p_*$ is its critical
value, such that data with $p>p_*$ form a black hole, while data with
$p<p_*$ disperse. The critical exponent $\gamma$ is independent of the
initial data, and depends only on the symmetry of the system under
consideration and type of matter. It can be related to the growth rate
of the unstable mode of the critical solution
\cite{livingreviews,KHA}.

Type II critical phenomena were first described by Choptuik
\cite{Choptuik} for the spherically symmetric massless scalar
field. Abrahams and Evans \cite{AbrahamsEvans} described type II
critical phenomena for axisymmetric vacuum gravity. In both cases the
critical solution is DSS. The critical collapse of a spherically
symmetric perfect fluid, the system which interests us here, was first
investigated by Evans and Coleman \cite{EvansColeman} and is CSS. Many
other systems were investigated subsequently and it seems that
critical phenomena are generic. For a recent review, see
\cite{livingreviews}.

Critical phenomena arise from any smooth initial data, in particular
from analytic initial data. The time evolution of these data is
analytic until the singularity occurs. (In fluids, it is possible that
a shock could form first, but empirically this is not the case.) The
critical solution, if it is to approximate such evolutions, must share
this analyticity property. A spherically symmetric CSS fluid critical
solution in particular must be analytic at the center to the past of
the singularity and at the sonic point to the past of the
singularity. These requirements define a boundary value problem
between the regular center and the first sonic point. This problem
admits a countable family of solutions which are parameterized by the
number of zeros of the radial velocity $V_R$. The critical solution
(with one unstable mode) turns out to be the one with a single zero
\cite{Harada}, with the fluid ingoing in a central region and outgoing
in the exterior. These solutions were first obtained numerically in
\cite{EvansColeman} for $\alpha=1/3$, in \cite{KHA,Maison} for
$0<\alpha<0.89$, and in \cite{NeilsenChoptuik} for $0<\alpha<1$.

Type II critical solutions have renewed interest in the study of
self-similar solutions in general relativity. We have seen that every
self-similar spacetime has a strong curvature singularity $s=0$
(unless it is flat). In critical solutions this singularity is naked,
in the sense that light rays can escape from regions of arbitrarily
high curvature with a finite redshift. In the collapse of
near-critical initial data, the solution is well-approximated by the
critical solution down to a lower cut-off scale of $s\sim
|p-p_*|$. Consequently the maximal value of the Kretschmann
scalar that can be seen is $|p-p_*|^{-4}$. In the limit in which the
parameter $p$ is tuned to its critical value $p_*$, the naked
singularity of the critical solution is seen. Therefore the naked
singularity of the critical solution is produced from a set of
codimension one in the space of smooth and asymptotically flat data.

Because the critical solution is determined by analyticity at the center in
the past of the singularity and the past sound cone of the
singularity, its continuation from the past sound cone is unique up to
the future light cone of the singularity, which acts as a Cauchy
horizon. We shall briefly discuss possible continuations beyond the
Cauchy horizon, but this is likely to be
irrelevant to gravitational collapse because quantum gravity effects
are expected to become important at very high curvature.

The nature of the critical solution in spherical perfect fluid
collapse with the equation of state $p=\alpha\mu$ has been discussed
by CCGNU \cite{CCGNU1}: For $0.28< \alpha<1$ the critical solution is
of the asymptotically Minkowski A type, and for the special value
$\alpha\simeq 0.28$ it is asymptotically Minkowski type B. In both
cases the spacetime structure is given by the left conformal diagram
in Fig.~\ref{fig:am1}.

The simplest continuation beyond the Cauchy horizon at $z=z_*$ would
be a wedge of flat empty spacetime, as indicated by the thin dashed
line in the left diagram in Fig.~\ref{fig:am1}. In this case, the CSS
singularity is a single point. The Ricci scalar at the center diverges
as $t^{-2}$ for $t<0$, and vanishes for $t>0$.

Another possible continuation would be the right conformal diagram in
Fig.~\ref{fig:am1}. From the general arguments in Section
\ref{section:dynamics} we see that such continuations form a
2-parameter family. In this case the singularity has a null branch
which is completely covered by a spacelike branch. Note that this
continuation is filled by fluid particles that emerge from the null
branch of the singularity and end at the spacelike branch. It could
therefore be described as a black hole expanding at the speed of light
or as a baby universe. A distant observer would see the $t^{-2}$
divergence of the curvature at the center, but would be swallowed by
the expanding black hole at $t=0$.

\begin{figure}
\includegraphics[width=7.5cm]{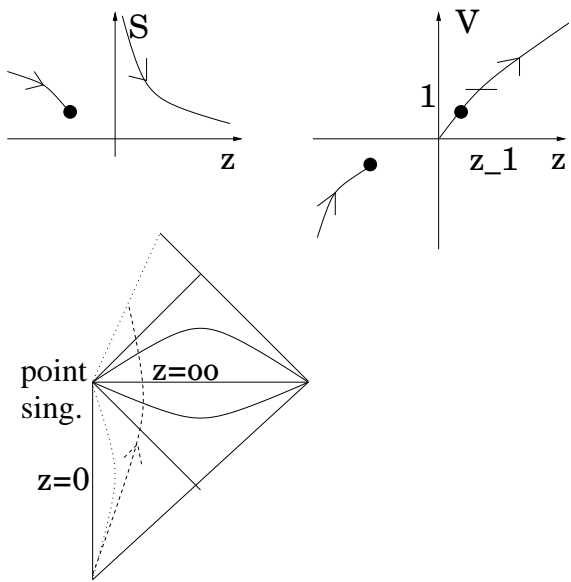}
\caption{The form of $S(z)$ and $V(z)$ and the conformal
diagram for the critical solutions with $0<\alpha < 0.28$. The
spacetime cannot be continued through the second sonic point as a CSS
and continuous solution, but it can be continued through a shock. The
two dashed lines are the sonic points, marked with dots
in the $S$ and $V$ plots.}
\label{fig:critical}
\end{figure}

For $0<\alpha < 0.28$, the critical solution is asymptotically
quasi-static, but its causal structure differs from the asymptotically
quasi-static solutions that we have discussed already. The reason is
that it crosses the sonic surface twice. The structure of the critical
solutions with $0<\alpha < 0.28$ up to the second sonic point is TST
and shown in Fig.~\ref{fig:critical}. We call the second sonic point
the sonic horizon, in analogy to the Cauchy horizon of the
singularity. CCGNU find that at the sonic horizon the critical
solution reaches the sonic surface away from the sonic line. As
discussed in Appendix~\ref{appendix:sonic}, this means that it cannot
be continued as a $C^0$ solution, so the density gradient and velocity
gradient become infinite at the sonic horizon.

This suggests that before the solution curve reaches the sonic surface
a shock occurs. It must be to the future of the Cauchy horizon and to
the past of the $z$-line where the CSS ansatz breaks down. Therefore
its location curve $r=r_{\rm sh}(t)$ is bounded by $z_{\rm ch}<r_{\rm
sh}(t)/t<z_{\rm sp}$. The most natural assumption compatible with this
is that the shock location is also self-similar, or $r_{\rm
sh}(t)/t=z_{\rm sh}={\rm const}$. Such solutions have been
investigated by Cahill and Taub \cite{CahillTaub}, and we need only
summarize their results here with our notation.

Given a CSS solution with a shock at $z_0$, the solution on one side
uniquely determines the solution on the other side. The Israel
matching conditions reduce in spherical symmetry to continuity of the
Hawking mass $m$ and the area radius $R$, so it follows that $M=m/R$
is continuous. We can also use the scale invariance of CSS solutions
to impose the gauge condition that the coordinate $r$ be
continuous. Continuity of $R$ is then equivalent to continuity of
$S$. The Rankine-Hugoniot conditions for the fluid give $V$ and
\begin{equation}
W\equiv \mu R^2= {\eta S^2\over 8\pi G}
\end{equation}
on one side in terms of their values on the other side (Eqs. (6.23-24)
of \cite{CahillTaub}). The complete matching conditions are
\begin{eqnarray}
S_+&=&S_-, \\
M_+&=&M_-, \\
V_+^2&=&{\alpha^2\over V_-^2}, \\
W_+&=&{(V_-^2-\alpha^2)W_-\over\alpha(1-V_-^2)} .
\end{eqnarray}
Note that $|V|<\sqrt{\alpha}$ on one side and $V>\sqrt{\alpha}$ on the
other. One also requires $|V|>\alpha$ on both sides for $W$ to be
positive. Therefore $V_-$ is restricted by the condition
$\alpha<|V_-|<\sqrt{\alpha}$. We thus have a 1-parameter family of
possible continuations parameterized by $V_-$ in this range, so the
existence of the shock adds one free parameter, namely $z_0$, just
like a nodal sonic point. In the strong shock limit $V_-^2\to\alpha^2$
we obtain $W_+\to0$ and $V_+^2\to1$, but the energy density measured
by a timelike observer, $\mu_+/(1-V_+^2)$, remains finite in this
limit. In the limit $V_-^2=\alpha$ the shock disappears, with
$W_+=W_-$ and $V_+=V_-$.


\section{Conclusions}


In this paper we have established the conformal structure of all
possible spherically symmetric CSS solutions, independently of the
Einstein equations. We have then
listed all possible conformal structures if the matter is a perfect
fluid with equation of state $p=\alpha\mu$. This includes the
solutions associated with critical phenomena in gravitational
collapse.

Our kinematical discussion is based on the ideas of Nolan \cite{Nolan}
and some of our diagrams were first drawn by Ori and Piran
\cite{OriPiran}. However, Nolan's kinematical discussion is restricted
to spacetimes compatible with ingoing null dust matter, and his use of
the Einstein equations is limited to imposing a positive energy
condition. Ori and Piran only consider spacetimes with a regular
center in the past.

The main limitation of our discussion is that we have excluded
solutions with shocks. Such solutions may have more free parameters
than differentiable ones and they may even admit additional conformal
diagrams.  Also we have only considered CSS continuations of the
critical solutions beyond their Cauchy horizon but this is probably
not a serious restriction because the continuation must be
asymptotically CSS at small scales.

By drawing the conformal diagrams we have been able to elucidate the
nature of some previously obscure CSS solutions. In particular, the
conformal structure of the asymptotically Minkowski solutions of type
A and type B are the same, and the CSS solutions that describe a black
hole forming in an open universe have a null singularity, which can
either be completely covered by the black hole singularity (see
Fig.~\ref{fig:am1} on the right) or covered only partly.

An interesting result is that, while the singularity may have past
branches from which fluid worldlines emerge and future branches where
they end, these branches must be connected by a null branch. All parts
of the singularity are therefore topologically connected. The
recollapsing models, for example, do not have the conformal diagram of
the closed Friedmann universe, with disjoint big bang and big crunch
singularities, but one of the conformal diagrams indicated in
Fig.~\ref{fig:af2}, Fig.~\ref{fig:af3} and Fig.~\ref{fig:aqs4}.

Reviewing earlier work \cite{CCGNU2}, we have placed the critical
solutions which arise in gravitational collapse within our general
classification. These solutions are analytic and unique up to the
Cauchy horizon of the CSS singularity. Intuitively one might expect
the continuation of the critical solution to be non-unique because it
depends on information coming out of the naked singularity. As
information in spherically symmetric fluid solutions travels only at
the speed of sound (because there are no gravitational waves), one
would expect the spacetime to be non-unique beyond the sonic horizon
of the singularity. However, contrary to this intuition, we have shown
that the critical solution is non-unique even before the sonic
horizon, at the Cauchy horizon.


\begin{acknowledgments}

We would like to thank J. M. Mart\'\i n-Garc\'\i a for interesting
discussions and for the use of his numerical code to compute the
critical solutions.

\end{acknowledgments}



\begin{appendix}



\section{The sonic surface}
\label{appendix:sonic}


Bicknell and Henriksen \cite{BicknellHenriksen} have expanded a
generic solution around a point in the sonic line in terms of a
parameter $u$ along the solution curve $(M,S,z,\eta)$. To linear order
they find
\begin{equation}
\label{zBH}
(\delta M,\delta S,\delta z,\delta \eta)(u) = \alpha_1 e^{\lambda_1 u}
{\bf V}_1 + \alpha_2 e^{\lambda_1 u} {\bf V}_2
\end{equation}
where the eigenvalues $\lambda_{1,2}$ and eigenvectors ${\bf V}_{1,2}$ are
known, and $\alpha_{1,2}$ are free constants. By parameter counting,
this linearized solution is generic. There are also two other
eigenvectors with zero eigenvalues in the 4-dimensional $(M,S,z,\eta)$
space. One takes the solution out of the constraint surface $c=0$, and
the other takes it along the sonic line. Their amplitudes can
consistently be set to zero.

The nature of the solution depends crucially on $\lambda_{1,2}$, which
are the solutions of a real quadratic equation. If they form a complex
conjugate pair the point on the sonic line is called a focus. The
formal solution (\ref{zBH}) is then complex and spirals around the
sonic point in phase space. No solution crosses the sonic surface at
this point.

If $\lambda_{1,2}$ are real with the same sign, the point on the sonic
line is called a node. We may assume without loss of generality that
$\lambda_1<\lambda_2<0$. Then the sonic line is crossed as $u\to
\infty$. Furthermore, as $u\to \infty$, $e^{\lambda_1 u}$ can be
neglected compared to $e^{\lambda_2 u}$, and the implicit functions
$M(z)$, $S(z)$, $\eta(z)$ are $C^1$. Therefore a 1-parameter family of
$C^1$ solutions crosses each point of the sonic line. All these
solutions have the same tangent at the sonic point. The solution with
$\alpha_1=0$ from this family is analytic. There is also an isolated
analytic solution with $\alpha_2=0$.

If $\lambda_{1,2}$ have opposite signs, the point on the sonic line is
called a saddle. Without loss of generality, we assume
$\lambda_1<0<\lambda_2$. Then there are only two isolated solutions
that cross the sonic surface: the solution with $\alpha_1=0$ as $u\to
-\infty$ and the solution with $\alpha_2=0$ as $u\to \infty$. Both
solutions are analytic.

The sonic line can be parameterized by $z$. It then turns out that the
points $0<z<z_1$ are saddles, $z_1<z<z_2$ are nodes, $z_2<z<z_3$ are
focal points, and $z_3<z<\infty$ are nodes again, where $z_{1,2,3}$
are known functions of $\alpha$ \cite{CarrColey}. The static solution
has a sonic point between $z_1$ and $z_2$, while the flat Friedmann
solution has a sonic point above $z_3$.

The two nodal segments of the sonic line are attractors, and are
crossed by generic 2-parameter families of solutions. The saddle
segment of the line is crossed only by a 1-parameter family of
solutions. Generic solutions therefore miss the saddle segment of the
sonic line and reach the sonic surface away from the sonic
line. Expanding around a point in the sonic surface but off the sonic
line, where $V^2=\alpha$ but $g\ne 0$, we find to leading order
that
\begin{eqnarray}
\delta M&\simeq& a\, \delta z, \\ 
\delta S&\simeq& b\, \delta z, \\
\delta \eta&\simeq&\pm\sqrt{2g
\left({\partial V^2\over \partial \eta}\right)^{-1}\,\delta z}, \\
\delta V&\simeq&\pm\sqrt{2g
{\partial V^2\over \partial \eta}\,\delta z}
\end{eqnarray}
This point on the sonic surface is reached by two solution curves from
one side, but there is no solution curve that reaches it from the
other side. Note also that, because $R=rS$ and $8\pi G\mu=r^{-2}\eta$,
the comoving physical density gradient $d\mu/dR$ and velocity gradient
$dV/dR$ diverge.


\section{Asymptotically Minkowski type B}
\label{appendix:MinkowskiB}


From Eq. (3.43) of \cite{CarrColey} we see that as $z\to\infty$, the
spacetime metric is approximately of the form
\begin{equation}
\label{A1}
ds^2 = - A^2 z^{2a}\,dt^2 + B^2 z^{2b} dr^2 + S_0^2 r^2
z^{2c}\,d\Omega^2
\end{equation}
where 
\begin{equation}
B/A=V_*, \quad a-1=b=1/(V_*^2-1), \quad c=1/(V_*-1).
\end{equation}
where $V_*$ is given by Eq. (\ref{V*}).
This (approximate) metric can be written as
\begin{equation}
\label{A3}
ds^2=-C\,du\,dv + S_0^2 \,v^2\,d\Omega^2
\end{equation}
where $C>0$ is a known constant, and 
\begin{equation}
u=-r z^{-1/(1+V_*)},\qquad v=r z^{1/(V_*-1)},
\end{equation}
are negative and positive, respectively.  Contrary to the claim in
\cite{CarrColey}, this metric is not Minkowski. From the form of the
metric, it is clear that $u$ and $v$ are affine parameters of null
geodesics. $z=\infty$ has two branches: $u=0$, which is therefore at
finite affine parameter distance, and $v=\infty$, which is at infinite
affine parameter distance. The former is a regular surface and the
spacetime can be continued there. The latter is $I^+$. Fluid worldlines
$r={\rm const}$ end up at $(u=0,v=\infty)$. Note that this point is
not $i_0$: observers that resist the outward fluid pressure can cross
the surface $u=0$.  It should be stressed that metrics (\ref{A1}) and
(\ref{A3}) are only approximate. Were they exact, the mass function
would be $m/R=1/2$ because $R=S_0v$. However, non-leading terms
actually show that $m/R\to 0$ in the limit $z\to \infty$.


\end{appendix}




\end{document}